\documentclass[aps,twocolumn]{revtex4}%
\usepackage{amsfonts}
\usepackage{amsmath}
\usepackage{amssymb}
\usepackage{graphicx}%
\providecommand{\U}[1]{\protect\rule{.1in}{.1in}}

\def\state#1{\mbox{$|#1\rangle$}}
\begin{document} 
\title{Dark-state polaritons for multi-component and stationary light fields}
\author{F.~E.~Zimmer}
\affiliation{\textit{SUPA}, School of Engineering and Physical Sciences, Heriot-Watt
University, Edinburgh, EH14 4AS, United Kingdom}
\author{J. Otterbach}
\affiliation{Fachbereich Physik der Technischen Universit{\"a}t Kaiserslautern, 67663
Kaiserslautern, Germany}
\author{R.G. Unanyan}
\affiliation{Fachbereich Physik der Technischen Universit{\"a}t Kaiserslautern, 67663
Kaiserslautern, Germany}
\author{B. W. Shore}
\affiliation{Fachbereich Physik der Technischen Universit{\"a}t Kaiserslautern, 67663
Kaiserslautern, Germany}
\author{M. Fleischhauer}
\affiliation{Fachbereich Physik der Technischen Universit{\"a}t Kaiserslautern, 67663
Kaiserslautern, Germany}
\date{May 16th, 2008}

\begin{abstract}
We present a general scheme to determine the loss-free adiabatic
eigensolutions (dark-state polaritons) of the interaction of multiple probe
laser beams with a coherently driven atomic ensemble under conditions of
electromagnetically induced transparency. To this end we generalize the
Morris-Shore transformation to linearized Heisenberg-Langevin equations
describing the coupled light-matter system in the weak excitation limit. 
For the simple lambda-type coupling scheme the generalized Morris-Shore transformation
reproduces the dark-state polariton solutions of slow light. Here we treat a closed-loop dual-$V$ scheme wherein two counter-propagating control fields generate a quasi stationary pattern of two
counter-propagating probe fields -- so-called stationary light. We show that 
 contrary to previous predictions,
there exists a single unique dark-state
polariton;  it obeys a simple propagation equation. 

\end{abstract}
\maketitle

\section{Introduction}
The adiabatic interaction of probe light pulses with coherently driven
three-level systems under conditions of electromagnetically induced
transparency (EIT) \cite{Fleischhauer-RevModPhys-2005} can most conveniently
be described in terms of so-called dark-state polaritons
\cite{Fleischhauer-PRL-2000}. These are the eigensolutions of
the Raman interaction of probe and control fields with the atomic ensemble in
the absence of atomic losses. They provide a simple theoretical framework for
phenomena such as slow light \cite{slow-light}, the storage and retrieval of
coherent light pulses in atomic ensemble \cite{Liu-Nature-2001,Phillips} and
quantum memories for photons
\cite{Fleischhauer-PRL-2000,Fleischhauer-PRA-2002}. Most importantly they
also fully
{incorporate} the essential physics of  adiabatic pulse propagation
  for time-varying control fields together with the associated coherent transfer of
excitation and quantum state from light to matter and vice versa. This transfer cannot be
understood in terms of electromagentic field equations alone. 

Recently it has been shown that the simultaneous presence of two counter-propagating control
fields of comparable strength can lead to a quasi-stationary pattern of slow
light consisting of two counter-propagating probe field components
\cite{Andre-PRL-2002,Bajcsy-Nature-2003,Moiseev-PRA-2006,Zimmer-OC-2006}.
Such stationary light pulses hold
particular interest as examples of  efficient nonlinear optical processes. In contrast
to the three-level coupling scheme, there exists no unique description of
stationary light in terms of dark-state polaritons. 
Reference  \cite{Moelmer,Moiseev-PRA-2006} introduced separate polariton fields for forward and
backward propagating modes. However, as we will show here,
these are not the quasi loss-less eigensolutions of the system.

We here determine the adiabatic eigensolutions of a coupled multi-level system
using a generalization of the stationary-light scheme. As with simpler systems, the
eigensolutions are superpositions of collective atomic operators and operators
of the electromagnetic field. Our procedure relies on the possibility of casting the
linearized Maxwell-Bloch equations into a matrix differential equation similar to that
used in treating the time-dependent Schr{\"o}dinger equation (TDSE) describing
coherent excitation of a  multi-level atom. For that equation there exists a well-known  
formalism to determine possible non-decaying eigenstates, the
Morris-Shore transformation \cite{Morris-PRA-1983}.
Generalizing this transformation to the linear Heisenberg-Langevin equations
of the coupled light-matter system we can easily determine their loss-free
eigensolutions, when they exist.
We apply this procedure to the stationary-light equations to find a dark-state polariton solution, and to derive its equation of propagation.


\section{The dual-$V$ stationary-light scheme}

\label{sec:TheModel}
%

\begin{figure}[t]
\includegraphics[width=0.45\textwidth]{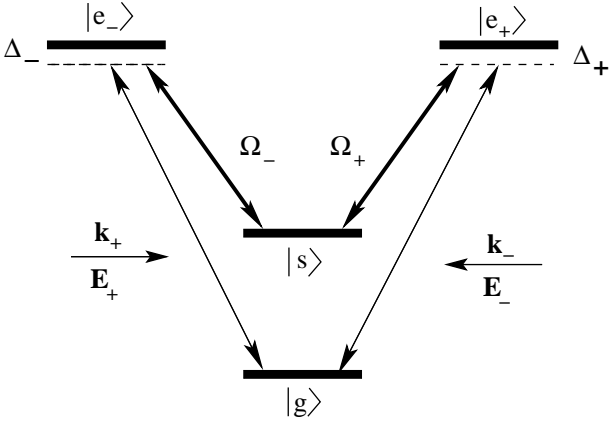}
\caption{The
dual-$V$ linkage pattern shown together with the depicted field configuration of
counter-propagating control ($\Omega_{\pm}$) and induced probe fields
($\hat{E}_{\pm}$) leads to stationary probe pulses. }%
\label{fig:2VCouplingScheme}%
\end{figure}

Let us consider one-dimensional propagation in the $z$ direction of two counterpropagating {\em probe fields}, circularly polarized orthogonally, having electric fields operators  
${\hat E}_{\pm}(z,t)$  and propagation vectors $k_{\pm}$. 
These interact with an ensemble of stationary four-state atoms via linkages between the ground state 
$\state{g}$ and excited states $\state{e_{\pm}}$. 
A second pair of  counterpropagating fields, also circularly polarized orthogonally and termed 
{\em control fields}, link these excited states to a fourth state 
$\state{s}$, through interactions characterized by Rabi frequencies $\widetilde\Omega_{\pm}(z,t) $ .
Figure  \ref{fig:2VCouplingScheme} illustrates the linkage pattern of the four fields with a four-state atom.
The control fields  are assumed to be sufficiently strong that they are not influenced by the atoms;  unlike the probe fields, they are not considered as dynamical variables. 

We assume that the control fields generate electromagnetically-induced transparency (EIT) for the probe
fields,  by requiring that their Rabi-frequencies are locked to those
of the control fields. Thus if $|\widetilde\Omega_{+}| =|\widetilde\Omega_{-}|$ the control
fields form a standing-wave pattern and the group velocity of the probe fields vanish; these form a standing-wave pattern known as \textit{stationary light}
\cite{Andre-PRL-2002,Bajcsy-Nature-2003}.

 A stationary-light probe-field pulse can
be created from a propagating-light probe field by adiabatically changing the
relative intensity of the two control fields from zero to unity, thereby  storing
the probe pulse in a collective spin excitation \cite{Zimmer-OC-2006}. The stored excitation can then be retrieved as a stationary pulse by using  two counter-propagating control fields of
equal intensity. 

Stationary-light pulses were suggested and experimentally demonstrated for the first time for a $\Lambda$-type three-state configuration by Lukin and co-workers \cite{Andre-PRL-2002,Bajcsy-Nature-2003}. In contrast to the system discussed here, there the counter-propagating fields, whether probe or control field, acted upon the same transitions.
A simple theoretical description of the $\Lambda$-type system of \cite{Bajcsy-Nature-2003}
is possible, by neglecting
components of the atomic spin coherence that are rapidly oscillating in space.
Such a secular approximation is only justifiable for a vapor at sufficiently
high temperature, such that  thermal motion of the atoms leads to a rapid
dephasing of the rapidly oscillating components
\cite{Moiseev-PRA-2006,Moelmer}. 
Here we avoid this dephasing approximation and consider 
the 4-level generalization shown
in Fig. \ref{fig:2VCouplingScheme}. This model  also produces
stationary light but does not require any secular approximation.

We assume that the control fields have carrier frequencies 
$\omega_{\pm}^{(c)}$ and wave-vectors 
$ k_{\pm}^{(c)} = \pm k^{(c)}\, $ directed along the $z$ axis.
We take the carrier frequencies and wave-vectors of the probe-field
components to be
$\omega_{\pm}^{(p)}$ and $  k_{\pm}^{(p)} = \pm k^{(p)}$, again along the $z$ axis. 
We replace the basic parameters of the control fields
${\widetilde\Omega}_{\pm}(z,t)$
and the operators $ \hat E_{\pm}(z,t) $ of the probe fields
by quantities
${\Omega}_{\pm}(z,t)
 \equiv {\widetilde\Omega}_{\pm}(z,t)
  \, \mathrm{e}^{-i (k_{\pm}^{(c)} z - \omega_{\pm}^{(c)}t)}$
   and  
   $\hat{\mathcal{E}}_{\pm}(z,t)
\equiv \hat E_{\pm}(z,t) \,
 \mathrm{e}^{-i (k_{\pm}^{(p)} z - \omega_{\pm}^{(p)}t)}$  
that vary  slowly with  $z$ and $t $. We further require four-photon
resonance and phase-matching, as expressed by the constraints
\begin{align}
\omega_{+}^{(c)}-\omega_{+}^{(p)}  & =\omega_{-}^{(c)}-\omega_{-}%
^{(p)},\nonumber\\
k_{+}^{(c)} - k_{+}^{(p)}  & = k_{-}^{(c)} - k_{-}^{(p)}%
.\label{eq:four-photon-resonance}%
\end{align}
Electromagnetically induced transparency requires
that the control and probe fields for both the forward and the backward
propagation direction are in two-photon resonance with the atomic system,
i.e.
\begin{align}
\omega_{+}^{(p)}-\omega_{+}^{(c)} = \omega_{-}^{(p)}-\omega_{-}^{(c)} =
\omega_{sg},
\end{align}
where $\omega_{sg}$ is the transition frequency between the ground state
$|g\rangle$ and the spin state $|s\rangle$.

To describe the atomic system we use collective atomic variables
${ {\widetilde\sigma}}_{\mu\nu}(z,t)$,   generalizations of
the traditional single-atom transition operators 
${\hat\sigma}_{\mu\nu} = |\mu\rangle \langle\nu|$ 
between states $|\nu\rangle$ and $|\mu\rangle$
\cite{Fleischhauer-PRL-2000,Fleischhauer-PRA-2002}. 
We take the two probe fields to have the same field-atom coupling strength, denoted $g$,
and we denote by $N$ the one-dimensional density of atoms, taken to be uniform. Then
the  interaction with
the light fields is described, in the rotating-wave approximation,  
by the interaction Hamiltonian
\begin{align}
\hat{H}^{\mathrm{int}}  &  
=\hbar\sum_{j=\pm}\int\mathrm{d}z \Bigl(
\omega_{e_{j}g}
{\hat\sigma}_{ e_{j} e_{j}}
- \Omega_{j} {\ e}^{i(k_{j}^{(c)} z
-\omega_{j}^{(c)}t)}\, {\widetilde\sigma}_{ e_{j} s}\nonumber\\
& \qquad\left. -g\sqrt{N} \hat{\mathcal{E}}_{j} \mathrm{e}^{i(k_{j}^{(p)}
z -\omega_{j}^{(p)}t)}\,   {\widetilde\sigma}_{ e_{j} g}
+\mathrm{h.~a.} \right] .\label{eq:InteractionHamiltonian-0}%
\end{align}
The
rapidly oscillating exponents in (\ref{eq:InteractionHamiltonian-0}) can be
eliminated by introducing slowly varying amplitudes of the atomic variables
\begin{align}
{\hat\sigma}_{ e_{\pm}s}(z,t)  & = { {\widetilde\sigma}}_{ e_{\pm}s}(z,t)\,
{\ e}^{i(k_{\pm}^{(c)} z -\omega_{\pm}^{(c)}t)},\\
{\hat\sigma}_{ e_{\pm}g}(z,t)  & = { {\widetilde\sigma}}_{ e_{\pm}g}(z,t)\,
{\ e}^{i(k_{\pm}^{(p)} z -\omega_{\pm}^{(p)}t)},\\
{\hat\sigma}_{ s g}(z,t)  & = { {\widetilde\sigma}}_{ s g}(z,t)\,
\mathrm{e}^{i((k_{\pm}^{(p)}-k_{\pm}^{(c)}) z -(\omega_{\pm}^{(p)}%
-\omega_{\pm}^{(p)})t)}.
\end{align}
In contrast to the original stationary-light scheme
\cite{Bajcsy-Nature-2003,Zimmer-OC-2006},  
the conditions
(\ref{eq:four-photon-resonance}) allow the introduction of 
  a slowly-varying amplitude of the ground state coherence 
  ${\hat \sigma}_{gs}$ without a secular approximation.   
The coherent interaction Hamiltonian can now be rewritten as
\begin{align}
\hat{H}^{\mathrm{int}}=-\hbar\sum_{j=\pm}\int\mathrm{d}z \Bigl[  
&
  \Delta_{j}\hat\sigma_{ e_{j} e_{j}} +\Omega_{j}\hat\sigma_{ e_{j} s}\nonumber
  \\
&  \left. +g\sqrt{N} \hat{\mathcal{E}}_{j}\hat\sigma_{ e_{j} g}+\mathrm{h.~a.} \right] .
\label{eq:InteractionHamiltonian}%
\end{align}
Here $\Delta_{\pm}\equiv\omega_{\pm}^{(p)}-\omega_{e_{\pm}g}$ are the
one-photon-detunings of the forward $(+)$ and backward $(-)$ propagating
probe fields respectively. 

The atomic system is also subject to losses and decoherence. These we treat 
by a set of Heisenberg-Langevin 
equations for the collective atomic operators and the operators for the probe
fields \cite{scully}. 
We assume that the control fields are sufficiently strong to
remain undepleted -- they are not dynamical variables -- and we
consider only the linear response of the atoms to the probe fields. 
In this limit  we treat the probe fields as perturbations in the Heisenberg-Langevin equations.

In zeroth order  all
atoms will be optically pumped into the ground state $|g\rangle$. The only nonzero atomic variable in this limit is   
${\hat\sigma}_{gg}^{(0)} = 1$;  other operators vanish, ${\hat\sigma
}_{\mu\nu}^{(0)}=0$. 

In first order one finds a linear set of
Heisenberg-Langevin equations for the atomic variables. These separate into
two uncoupled sets. Those relevant for this work are
\begin{align}
\partial_{t}\hat{\sigma}_{ ge_{\pm}}= & -\left[ {\ i}\Delta_{\pm}
+\gamma_{ge_{\pm}}\right] \hat{\sigma}_{ ge_{\pm}}\nonumber\\
& +\mathrm{i} g \sqrt{N} \hat{\mathcal{E}}_{\pm}+\mathrm{i}\Omega_{\pm}%
\hat\sigma_{ gs}+ \hat{F}_{ ge_{\pm}}%
,\label{eqn:HeisenbergLangevinWeakProbeField1}\\
\partial_{t}\hat{\sigma}_{ gs}= &  +\mathrm{i} \Omega^{*}_{+} \hat\sigma_{
ge_{+}}+{\ i} \Omega^{*}_{-}\hat\sigma_{ ge_{-}},%
 \label{eqn:HeisenbergLangevinWeakProbeField2}%
\end{align}
where the quantities $\gamma_{ge\pm}$ parametrize the losses and decoherence rates
and the $\hat{F}_{ge\pm}$ are Langevin noise forces \cite{scully,gardiner}
associated with the decays. 
In eq.~(\ref{eqn:HeisenbergLangevinWeakProbeField2})
we have assumed that the ground-state coherence ${\hat\sigma}_{ gs}$ is 
stable. This is appropriate for hyperfine-states in cold and hot atomic vapours 
for which the coherence lifetimes, mainly caused by elastic collisions, are of the order of milliseconds
\cite{Liu-Nature-2001,Phillips,Bajcsy-Nature-2003}. As a consequence of this stability the equation includes no corresponding Langevin noise operator.

The Langevin noise operators $ \hat F_{A}(t)$ are necessary to preserve the
commutation relations of quantum variables. We assume, as is customary, that
the associated decay is exponential in time, so that the noise operators are 
delta-correlated in time, 
\begin{equation}
\langle \hat F_{A}(t) \hat F_{B}(t^{\prime})\rangle= D_{AB}(t)\, \delta(t-t^{\prime
}).\label{eq:noise}
\end{equation}
 The  diffusion coefficient  $D_{AB}(t)$  appearing here is  related to the
dissipative part of the dynamics through the dissipation-fluctuation
theorem \cite{scully,gardiner} 
\begin{eqnarray*}
 D_{AB}(t) &=& \frac{{\rm d}}{{\rm d}t} \Bigl\langle \hat A(t) \hat B(t)\Bigr\rangle_{\rm loss}
-\Bigl\langle \frac{{\rm d}}{{\rm d}t}\hat A(t)\Bigr\vert_{\rm loss}\!\! \hat B(t)\Bigr\rangle\nonumber\\
&&-\Bigl\langle \hat A(t)\frac{{\rm d}}{{\rm d}t}\hat B(t)\Bigr\vert_{\rm loss}\Bigr\rangle.
\end{eqnarray*}
 One easily verifies that, 
 because the quantum noise
originates from spontaneous emission processes,
   the relevant diffusion coefficients of the present system
are proportional to the excited-state populations.  However, in the linear response limit considered here 
this population is negligible and thus we can ignore the Langevin noise terms altogether. 

To complete the description of the atom-field system we require equations for the probe fields. 
These we take to be wave equations for the slowly-varying probe
field amplitudes
\begin{equation}
\left[ \partial_{t}\pm c\partial_{z} \right] \hat{\mathcal{E}}_{\pm}=i g
\sqrt{N}\hat\sigma_{ ge_{\pm}}. \label{eq:ShortenedWaveEquation}%
\end{equation}
These, together with the atomic equations
(\ref{eqn:HeisenbergLangevinWeakProbeField1}) and
(\ref{eqn:HeisenbergLangevinWeakProbeField2}) 
form  the set of 
  self-consistent set of Maxwell-Bloch equations  which are the basis of the following considerations.

We note that equations (\ref{eqn:HeisenbergLangevinWeakProbeField1}) 
and (\ref{eqn:HeisenbergLangevinWeakProbeField2}) 
are formally identical to the time-dependent
Schr{\"o}dinger equation (or TDSE) for the $\Lambda$-system after the secular
approximation (see e.g. eq.~(9) and (10) in \cite{Zimmer-OC-2006}).  No such
approximation has been used here, however. Furthermore, in contrast to the
system studied in \cite{Bajcsy-Nature-2003,Zimmer-OC-2006} the single-photon
detunings $\Delta_{\pm}$ can be chosen to be different for the forward and
backward direction, which adds another degree of control. 

  We now assume that the
slowly varying amplitudes of the control field Rabi-frequencies are constant
in space, as is  the atom density $N$. We introduce  a spatial Fourier transformation, through variables
 $X(z,t) = \int\mathrm{d}k\, \exp(-i k z)\, X(k,t)$, thereby reducing the 
Heisenberg equations and the
shortened wave equation to coupled
differential equations in time alone.
We write these in matrix form as
\begin{equation}
\frac{d}{d t} \mbox{\bf X}  = - i \mathsf{H} \mbox{\bf X} 
\label{eqn-X}
\end{equation}
where the column vector of dynamical variables has the elements 
$\mbox{\bf X}^{\top}
 = \{ \hat{\mathcal{E}}_{+},\hat{\mathcal{E}}_{-}%
 ,\hat\sigma_{ gs},\hat\sigma_{ ge_{+}},\hat\sigma_{ ge_{-}} \}
$,
 and the  matrix of (slowly varying) coefficients is
\begin{align}
\mathsf{H} = \left[
\begin{matrix}
 - kc & 0 & 0 & -g \sqrt{N} & 0\\
0 & +kc & 0 & 0 & -g \sqrt{N}\\
0 & 0 & 0 & -\Omega_{+} & -\Omega_{-}\\
-g\sqrt{N} & 0 & -\Omega_{+}^{*} & -i\Gamma_{ge_{+}} & 0\\
0 & -g \sqrt{N} & -\Omega_{-}^{*} & 0 & - i\Gamma_{ge_{-}}%
\end{matrix}
\right] .\label{eq:schrodinger-W}%
\end{align}
To simplify typography we have introduced the symbol
\begin{equation}
 \Gamma_{ge_{\pm}} \equiv \mathrm{i}\Delta_{\pm}+\gamma_{ ge_{\pm}}.
\end{equation}
The linkage pattern  of these equations is shown in Fig. \ref{fig:M-scheme}. It
corresponds to the M-type linkage found with the TDSE for light fields interacting with
atoms, as was extensivly studied in the
literature \cite{shore}. In the following we draw on this analogy with the conventional TDSE  to  identify the loss-free adiabatic
eigensolutions of the system through use of the 
Morris-Shore transformation, a method widely used to find adiabatic dark
states of the interaction of atomic multi-level systems with near-resonant
laser fields.


\section{The Morris-Shore transformation}

\label{sec:TheMorrisShoreTrafo}
%

 Let us first 
%
summarize the basic properties of the Morris-Shore transformation
\cite{Morris-PRA-1983}, before
considering the two examples most relevant for us.

 Consider a first-order ODE, of the form (\ref{eqn-X}), in which 
 $\mbox{\bf X}$ is an
 $N$-dimensional column vector that separates into two classes
of variables: a set $A$ of $N_{A}$ variables and a set $B$ of $N_{B}$
variables. We suppose that the $N \times N$ matrix $\mathsf{H} $ has elements only between
the $A$ and $B$ sets, not within them. $\mbox{\bf X} $ and $\mathsf{H}$  therefore have the forms
\begin{equation}
\mbox{\bf X} = \left[
\begin{array}
[c]{c}%
\mbox{\bf X}_{A}\\
\mbox{\bf X}_{B}%
\end{array}
\right] , 
\qquad \mathsf{H} = \left[
\begin{array}
[c]{cc}%
\mbox{\sf 0} & \mathsf{V}\\
\mathsf{V}^{\dag} & \mbox{\sf 0}
\end{array}
\right] ,
\end{equation}
 where $\mathsf{V}$ is an $N_{A}\times N_{B}$ matrix. 
%
We further assume that initially (at $t = 0$) all variables of the $B$ set vanish,  
$\mbox{\bf X}_{B}(t=0) = 0$. Such equations can, by means of 
  the Morris-Shore (MS) transformation 
$\mbox{\bf Y} = \mathsf{M}^{\dag} \mbox{\bf X} \mathsf{M}$
\cite{Morris-PRA-1983}, be rewritten as a
set of independent equations involving only pairs of variables (one being a
combination of the $A$ set variables, the other a combination from the $B$
set) or unlinked single variables. The number of unlinked (``dark'') variables
$N_{D}$ is the difference $N_{D} = |N_{A} - N_{B}|$. In the new basis the
equation of motion reads
\begin{equation}
\frac{d}{d t} \mbox{\bf Y} = -i \mathsf{H} ^{MS} \mbox{\bf Y}
\end{equation}
where the matrix of coefficients has the structure
\begin{equation}
\mathsf{H} ^{MS} \equiv\mathsf{M}^{\dag} \mathsf{H} \mathsf{M} = \left[
\begin{array}
[c]{ccccc}%
\mbox{\sf w}^{(1)} & {\mbox{\bf 0}} & {\mbox{\bf 0}} & \cdots & \mbox{\bf 0}\\
{\mbox{\bf 0}} & \mbox{\sf w}^{(2)} & {\mbox{\bf 0}} & \cdots & \mbox{\bf 0}\\
{\mbox{\bf 0}} & {\mbox{\bf 0}} & \mbox{\sf w}^{(3)} & \cdots & \mbox{\bf 0}\\
\vdots & \vdots & \vdots & \ddots & \mbox{\bf 0}\\
\mbox{\bf 0} & \mbox{\bf 0} & \mbox{\bf 0} & \mbox{\bf 0} & \mbox{\bf 0}
\end{array}
\right] .
\end{equation}
%
Here the $\mbox{\sf w}^{(i)}$ are matrices of dimension $2 \times 2$
\begin{equation}
\mbox{\sf w}^{(j)} = \left[
\begin{array}
[c]{cc}%
0 & v^{(j)}\\
v^{(j)} & 0
\end{array}
\right].
\end{equation}
There are  min($N_{A}, N_{B}$) of these, each  linking pairs of ``bright'' variables. The remaining 
``dark" variables
 remain
  constant, i.e.
\begin{equation}
\frac{d}{d t} Y_{n} = 0.
\end{equation}

For the  $\Lambda$ and dual-$V$ linkages
 considered here the $A$ set includes variables representing both field
and atom degrees of freedom. The resulting MS dark variables, combining field and atom properties, are  dark-state polaritons.

\subsection{The $\Lambda$-system}


For the $\Lambda$-system ($N = 3$ with $N_{A} = 2, N_{B} = 1$ and hence $N_{D}
= 1$ dark state) the original matrix, when expressed in a basis   $\{X_1,X_2,X_3\}$ , has 
the form (apart from reordering)
\begin{equation}
\left[
\begin{array}
[c]{ccc}%
0 & 0 & V_{1}\\
0 & 0 & V_{2}\\
V_{1} & V_{2} & 0
\end{array}
\right]. \label{eq:Lambda-matrix}%
\end{equation}
The single dark variable, constructed entirely from $A$ variables, is
\begin{equation}
Y_{D}  = \left[  V_{2} X_{1} - V_{1} X_{2}  \right]  / \mathcal{N}%
\label{eq:LambdaDarkState}%
\end{equation}
where $\mathcal{N}$ is a normalization factor. Note that
$Y_{D}$ remains a ``dark" variable even when   a finite diagonal element is added
to the third row of the matrix in (\ref{eq:Lambda-matrix}). 

To connect this formalism with that of dark-state polaritons we let $\mbox{\bf X}$
represent  the amplitudes of the wavevector of the atomic $\Lambda$ system,
with $X_{1}$ and $X_{2}$ being the amplitudes of the stable (i.e.
non-decaying) lower levels. Then $Y_{D}$ is the well-known dark state of that system.


\subsection{The M-system}


The M-system ($N = 5$ with $N_{A} = 3, N_{B} = 2$ ) has also only one dark
variable. The original matrix has the form (apart from reordering)
\begin{equation}
\left[
\begin{array}
[c]{ccccc}%
0 & 0 & 0 & V_{1} & 0\\
0 & 0 & 0 & 0 & V_{2}\\
0 & 0 & 0 & V_{3} & V_{4}\\
V_{1} & 0 & V_{3} & 0 & 0\\
0 & V_{2} & V_{4} & 0 & 0
\end{array}
\right] .
\end{equation}
The single dark variable is
\begin{equation}
Y_{D} = \left[  V_{2} V_{3} X_{1}  +V_{1} V_{4} X_{2}  - V_{1} V_{2} X_{3}
\right]  / \mathcal{N}\label{eq:dark-variable-2V}%
\end{equation}
where $\mathcal{N}$ is a normalization constant. As with the
$\Lambda$-system, $Y_{D}$ remains a ``dark" state if diagonal elements are added
to the  two lowest rows of the matrix.


\section{Application of the Morris-Shore transformation to the dual-$V$ system}
\label{sec:Application}

\begin{figure}[t]
\includegraphics[width=0.45\textwidth]{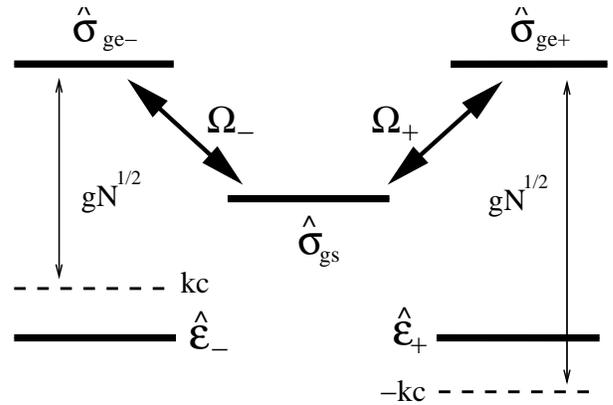}
\caption{The
$M$-scheme corresponding to the dual-$V$ coupling scheme 
depicted in Fig. \ref{fig:2VCouplingScheme}. The matter and field amplitudes are treated as
basis states of the new system.}%
\label{fig:M-scheme}%
\end{figure}

We now apply this 
mathematical tool to actual physical problems. The dual-$V$ system depicted in Fig.\ref{fig:2VCouplingScheme} can
be described  in the Heisenberg picture  as the M-system shown in
Fig. \ref{fig:M-scheme}. 
Applying the Morris-Shore transformation one can
immediatly construct the dark variable. Neglecting the ($\pm kc$) diagonal
elements we find
\begin{equation}
Y_{D}=\left[ \Omega_{+}g\sqrt{N}\hat{\mathcal{E}}_{+}+\Omega_{-}g\sqrt{N}\hat{\mathcal{E}}_{-}-g^{2}N\hat{\sigma
}_{gs}\right]  /\mathcal{N}.
\end{equation}
To express this more simply we define
  mixing angles by means of the relationships $\tan^{2}\theta={g^{2}N}/{\Omega^{2}}$ and
$\tan^{2}\varphi={\Omega_{-}^{2}}/{\Omega_{+}^{2}}$, where $\Omega
^{2}=\Omega_{+}^{2}+\Omega_{-}^{2}$. The construction then reads
\begin{equation}
Y_{D}  
=\left(  \cos\varphi\hat{\mathcal{E}}_{+}
+\sin\varphi\hat{\mathcal{E}}_{-}\right)
\cos\theta
  - \hat{\sigma}_{gs} \sin\theta.
\end{equation}
This is the Fourier transform of the dark-state polariton (DSP) for a dual-$V$-system. 
To obtain the adiabatic limit we did set $kc\rightarrow0$ in
eqs. \eqref{eq:schrodinger-W}, thereby neglecting 
all spatial variations; that limit produces the equation  
\begin{equation}
\frac{d}{d t}
Y_{D} = 0.
\label{eq:AdiabaticLimit}%
\end{equation}
In addition to the usual adiabatic condition $\Omega_{\text{eff}}T\gg1$, where
$\Omega_{\text{eff}}=\sqrt{g^{2}N+\Omega^{2}}$ and $T$ characterizes the pulse
duration, one must impose a further adiabatic condition on the spatial Fourier
frequencies to justify the negelct of the terms $\pm kc$.
As has been discussed in detail in
Ref.\cite{Fleischhauer-PRA-2002}
the corresponding condition is
\begin{equation}
L_{p}\gg\sqrt{L_{\text{abs}}}\sqrt{\frac{c}{\gamma}},
\end{equation}
where $L_{\text{abs}}= c\gamma / g^{2}N $ is the absorption length of the
medium, and $L_{p}$ is the characteristic length scale over which changes occur  in the amplitude
of the dark-state polariton in the medium. 

The adiabatic limit, eq. (\ref{eq:AdiabaticLimit}), does not account properly for
propagation effects  \cite{Vitanov},  e.g. it neglects consequences of nonzero group velocity.  
In order to describe the propagation
dynamics of the DSPs one needs at least the  lowest order corrections  to the spatial Fourier components.
These can be calculated in a very simple perturbative way. The introduction of  nonzero $kc\neq0$ amounts to adding the terms $\{ -kc,+kc,0,0,0 \}$ to the diagonal elements of $\mathsf{H}$, 
leading to the equation
\begin{equation}
\frac{d}{d t}\mbox{\bf Y}=-i\mathsf{H}^{MS}%
\mbox{\bf Y}-i\mathsf{M}^{\dagger}\left[
\begin{matrix}
-kc & 0 & 0 & 0 & 0\\
0 & +kc & 0 & 0 & 0\\
0 & 0 & 0 & 0 & 0\\
0 & 0 & 0 & 0 & 0\\
0 & 0 & 0 & 0 & 0
\end{matrix}
\right]  \mathsf{M}\mbox{\bf Y}.
\end{equation}
From this we deduce the dispersion relation for the dark-state polariton 
\begin{equation}
 \omega\left(  k\right)  =k\,C_1+k^2\,C_2.
\label{Dispersion}
\end{equation}
The coefficients appearing here are obtained as the
 first two non-vanishing corrections to the eigenvalue of
the dark-state polariton due to the terms proportional to $\pm k c$. 
\begin{subequations}
\begin{align}
 C_1 &= v_{gr} \cos2\varphi\\
 C_2 &=v_{gr} L_{\text{abs}}\left(\frac{\Delta}{\gamma}-i\right)\left(  \sin^2 2\varphi+\cos^{2}2\varphi\sin^4\theta\right). 
\end{align}
\end{subequations}
Here $v_{gr}=c\cos^2\theta$ is the group velocity of standard slow light \cite{Fleischhauer-PRL-2000} 
and we have assumed that
$\Gamma_{ge_{+}}=\Gamma_{ge_{-}}=\Gamma=\gamma+i\Delta$. 
 We see that the effective velocity of the dark-state polariton in the stationary light case is then given by 
\begin{equation}
v=\frac{d\omega}{dk}=c\cos^{2}\theta\cos2\varphi
\label{eqn-vg}
\end{equation}
 and that  the DSP  gains an effective  complex-valued  mass 
\begin{eqnarray}
\frac{1}{m^*} &\equiv& \frac{1}{\hbar} \frac{d^{2}\omega}{dk^{2}} \label{complexmass}\\
&=& \frac{4\pi}{m}\frac{v_{gr}}{v_{rec}}\frac{L_{\rm abs}}{\lambda} \left(\frac{\Delta}{\gamma}-i\right)%
\left(  \sin^2 2\varphi+\cos^{2}2\varphi\sin^4\theta\right).\nonumber
\end{eqnarray}
Here
 $\lambda=2\pi/k^{(p)}$ is  the wavelength of the probe field, $m$ is the atomic mass, and
$m v_{rec} = \hbar k^{(p)} $ the recoil momentum of a probe-field photon. In general the effective mass of the DSPs of stationary light is 
much smaller than the real mass of the atoms, even if the group velocity $v_{gr}$ is of the order of the recoil velocity $v_{rec}$, because
the wavelength $\lambda$ is much smaller than the absorption length $L_{\rm abs}$.

From the above considerations we obtain a very simple equation of motion for the dark-state
polaritons including the  lowest order corrections. The equation for Fourier components reads
\begin{equation}
\left[  \frac{d}{d t}+i k v
- i\frac{\hbar}{2 m^*} k^{2}\right] Y_{D}=0.
\end{equation}
Transformed to real space this equation becomes
\begin{equation}
\left[  \frac{\partial}{\partial t}-v\frac{\partial
}{\partial z}+i\frac{\hbar}{2 m^*}\frac{\partial^{2}}{\partial z^{2}}\right]  \hat{\Psi}_{D}=0.
\end{equation}
The effective group velocity
of the DSP, eq. (\ref{eqn-vg}), is  fully determind by the mixing angles. The DSP comes to rest, $v = 0$,   when at least one of
the two trigonometric functions, $\cos\theta$ or $\cos 2\varphi$, vanishes.  It is interesting to note
that the sign of $v$ can be changed by altering the mixing angle $\varphi$, i.e.
by changing the ratio of the Rabi frequencies of the two counter-propagating control
fields. If both these fields have the same strength  then $v=0$ and the field
pattern has no net motion. If the forward (backward) propagating control field
is stronger than the other one, the net motion of the
probe field pattern is also in the forward (backward) direction. 

One also notices that when  $\cos2\varphi=0$ the first-order term $C_1$ vanishes, and thus the dynamics is dominated by either a 
Schr\"{o}dinger-type evolution or a weak diffusive spreading of the probe field, depending on  
the relative size of imaginary and real parts of $m^*$. On the other hand, if
only one of the two control lasers is present then $\cos 2\varphi=\pm 1$ and we
recover the equation of motion for the slow light polaritons found in 
\cite{Fleischhauer-PRL-2000}.


It should also be noted that as soon as one knows the form of the dark state
one is able to calculate all other states of this system, also termed
bright-state polaritons. This can  be done by defining mutually orthogonal
eigensolutions,  starting with the dark state. One is able also to derive  
all non-adiabatic couplings arising e.g. from time dependent mixing angles, leading to first order corrections  to  the adiabatic approximation.

Let us finally address some  limiting factors of the considered system/approach in particular phase fluctuations of the control 
fields, collisional dephasing of the ground-state transition and Doppler broadening due to atomic motion.
In the above discussion we have disregarded processes which destroy either the amplitude or the phase of the
ground-state coherence $\hat\sigma_{gs}$. While the amplitude of ground-state superpositions is usually quite
robust, its phase can be destroyed by two processes: atomic collisions and fast phase fluctuations of the coupling
laser \cite{Dalton-JOSAB-1982}. If the laser phase fluctuations are sufficiently slow, such that adiabaticity 
holds true, probe and control fields are phase correlated and their difference phase does not
change \cite{Fleischhauer-PRL-1994}. Phase destroying collisions as well as fast laser phase fluctuations lead to
a decay of dark-state polaritons into bright polaritons which are subsequently absorbed. Thus the concept of the
dark polaritons is only useful if the latter process is sufficiently slow on the time scales of interest.

Let us secondly introduce the Doppler width $\Delta_D$  of the $|g\rangle-|e_\pm\rangle$-transitions and the factor $r_\pm=(\omega_{e\pm g}-\omega_{e\pm s})/\omega_{e\pm g}$.
Using these we note, that if the following conditions: $\Omega^2>>\gamma_{gs}\Delta_D$, $\Omega^2>>\Delta_D r_\pm$ and $\Omega^2>>(\gamma_{ge\pm}+\gamma_{gs\pm})\Delta_D r_\pm$ are met, the influence of Doppler broadening onto the absorptive and dispersive properties of a $\Lambda$-type three-level system is negligible as long as  the Doppler-free configuration of co-propagating control and probe fields is used \cite{Fleischhauer-PRA-1994}.  The dual-V stationary light scheme we have introduced here is essentially a double $\Lambda$-scheme with the two linkages coupling to the same coherence $\hat\sigma_{gs}$. Both $\Lambda$-type sub-systems work in the necessary Doppler-free configuration.  If the lower states $|g\rangle$ and $|s\rangle$ are hyperfine levels, the above conditions are practically always fulfilled. Hence, Doppler broadening is not a limiting factor here.  

FEZ thanks  P. \"Ohberg for discussions and acknowledges
financial support from the DFG graduate school ``Ultrakurzzeitphysik und
nichtlineare Optik" at the Technical University of Kaiserslautern as well as
the UK EPSRC. BWS acknowledges support from funds available through the Max Planck Forschungspreis 2003.





\begin{thebibliography}{99}                                                                                               %
\bibitem {Fleischhauer-RevModPhys-2005}M.~Fleischhauer, A.~Imamo\u{g}lu, and
J.P. Marangos. \newblock Electromagnetically induced transparency: Optics in
coherent media. \newblock {\em Rev. Mod. Phys.} {\bf 77}, 633--673 (2005).

\bibitem {Fleischhauer-PRL-2000}M.~Fleischhauer and M.~D. Lukin.
\newblock Dark-state polaritons in electromagnetically induced transparency.
\newblock
\emph{Phys. Rev. Lett.} \textbf{84}, 5094--5097 (2000).

\bibitem {slow-light}L. V. Hau, S. E. Harris, Z. Dutton and C. H. Behroozi.
\newblock Light speed reduction to 17 metres per second in an ultracold atomic
gas. \newblock{\em Nature} {\bf 397}, 594--598 (1999).

\bibitem {Liu-Nature-2001}C. Liu, Z. Dutton, C.H. Behroozi and L.V. Hau.
\newblock Observation of coherent optical information storage in an atomic
medium using halted light pulses. \newblock {\em Nature} {\bf 409}, 490--493 (2001).

\bibitem {Phillips}D.F. Phillips, A. Fleischhauer, A. Mair, R.L. Walsworth and
M.D. Lukin. \newblock Storage of Light in Atomic Vapor. \newblock {\em
Phys. Rev. Lett.} {\bf 86}, 783--786 (2001).

\bibitem {Fleischhauer-PRA-2002}M.~Fleischhauer and M.~D. Lukin.
\newblock Quantum memory for photons: Dark-state polaritons.
\newblock {\em Phys. Rev.
A (Atomic, Molecular, and Optical Physics)} \textbf{65}, 022314 (2002).

\bibitem {Andre-PRL-2002}A.~Andr{\'e} and M.~D. Lukin. \newblock Manipulating
light pulses via dynamically controlled photonic band gap.
\newblock {\em Phys. Rev. Lett.} {\bf 89}, 143602 (2002).

\bibitem {Moiseev-PRA-2006}S.A. Moiseev and B.S. Ham. \newblock Quantum
manipulation of two-color stationary light: Quantum wavelength conversion.
\newblock {\em Phys. Rev. A} {\bf 73}, 033812 (2006).

\bibitem {Bajcsy-Nature-2003}M.~Bajcsy, A.~S. Zibrov, and M.~D. Lukin.
\newblock Stationary pulses of light in an atomic medium. \newblock {\em
Nature} {\bf 426}, 638--641 (2003).

\bibitem {Zimmer-OC-2006}F.~E. Zimmer, A.~Andr{\'e}, M.~D. Lukin, and
M.~Fleischhauer. \newblock Coherent control of stationary light pulses.
\newblock {\em Optics Comm.} {\bf 264}, 441--453 (2006).

\bibitem {Moelmer}K.~Hansen and K.~Molmer. \newblock Stationary light pulses
in ultra cold atomic gasses. \newblock {\em Phys. Rev. A} {\bf 75}, 065804 (2007).

\bibitem {Morris-PRA-1983}J.~R. Morris and B.~W. Shore. \newblock Reduction of
degenerate two-level excitation to independent two-state systems.
\newblock {\em Phys. Rev. A} {\bf 27}, 906--912 (1983).

\bibitem {Vitanov}N.V. Vitanov, M. Fleischhauer, B.W. Shore and K. Bergmann.
\newblock Coherent manipulation of atoms and molecules by sequential laser
pulses. \newblock {\em Adv. Atom. Mol. Opt. Phys.} {\bf 46}, 55 (2001).

\bibitem {shore}B.W. Shore. \newblock {\em Theory of coherent atomic
excitation.} \newblock John Wiley \& Sons, New York, 1990.

\bibitem {scully}M.O. Scully and M.S. Zubairy. \newblock {\em Quantum
Optics.} \newblock Cambridge U.P., New York, 1997.

\bibitem {gardiner}C. W. Gardiner, \emph{Quantum Noise}, Springer, Heidelberg, 1991.

\bibitem{Dalton-JOSAB-1982}B.J. Dalton and P.L. Knight,\newblock The effects of laser field fluctuations on coherent population trapping.\newblock{\em J. Phys. B: At. Mol. Phys.} {\bf 15}, 3997 
(1982).

\bibitem{Fleischhauer-PRL-1994}M. Fleischhauer,\newblock Correlations of high-frequency phase fluctuations in electromagnetically
induced transparency.\newblock{\em Phys. Rev. Lett.} {\bf 72}, 989 (1994).

\bibitem{Fleischhauer-PRA-1994}M. Fleischhauer and M.O. Scully,\newblock Quantum sensitivity limits of an optical magnetometer based on atomic phase coherence.\newblock{\em Phys. Rev. A} {\bf 49}, 1973 (1994).

\end{thebibliography}

\end{document}